\documentclass[12pt,titlepage]{article}
\usepackage[english]{babel}
\usepackage[dvips]{graphicx}
\usepackage[dvips]{feynmf}
\usepackage{amsmath,amssymb}

\begin{document}

\begin{titlepage}
\begin{flushright}

\end{flushright}

\vspace{.5cm}
\begin{center}
{\Large \bf 
 The $B^- \to \phi \phi K^-$ decay rate with $\phi \phi$ invariant mass 
 below  charm threshold }

\vspace{.5cm}

{\large \bf S. Fajfer$^{a,b}$, T.N. Pham$^{c}$, A. Prapotnik$^{b}$\\}

\vspace{0.5cm}
{\it a) Department of Physics, University of Ljubljana, 
Jadranska 19, 1000 Ljubljana, Slovenia}

{\it b) J. Stefan Institute, Jamova 39, P. O. Box 300, 1001 Ljubljana, 
Slovenia}\vspace{.5cm}

{\it c) Centre de Physique Teorique, Centre National de la Recherche 
Scientifique, UMR 7644, Ecole Polytechnique, 91128 Palaiseau Cedex, France}
\end{center}
\centerline{\large \bf ABSTRACT}

\vspace{0.5cm}
We investigate the decay mechanism in the $B^- \to \phi \phi K^-$ 
decay with the $\phi\phi$ invariant mass below
the charm threshold and in the neighborhood of the  $\eta_c$ 
invariant mass region. Our approach is based on the use 
of factorization model and the knowledge of matrix elements of the weak 
currents. For the $B$ meson weak transition we apply 
form factor formalism, while for the light mesons we use effective weak and 
strong Lagrangians. { We find that the dominant contributions 
to the branching ratio come from the   $\eta$, $\eta'$ and 
$\eta(1490))$ pole 
terms of  the penguin operators  in the decay chains
 $B^- \to \eta (\eta', \eta(1490) ) K^- \to \phi \phi K^-$.}
Our prediction for the branching ratio  is in agreement with  the Belle's 
result.  

\end{titlepage}

{\bf I. INTRODUCTION}
\vspace{.5cm}

It is a very fruitful era in $B$ meson physics.  A lot of experimental 
data on $B$ meson decays  is coming from the $B$ meson factories. 
Many of their results are still not  explained. Recently,  
Belle collaboration has announced the observation of the 
${\rm BR}(B^{\pm} \to \phi \phi K^{\pm})=$ 
$(2.6^{+1.1}_{-0.9}\pm 0.3)\times 10^{-6}$ \cite{Belle} for a 
$\phi \phi$ invariant mass below $2.85\,$GeV.
This is the first of the three-body $B$ decays with  
two vector mesons and one pseudoscalar meson in the final state that 
has been observed. The $B$ meson decays into three pseudoscalar 
mesons have been studied \cite{cheng,BFOPP} within heavy quark 
symmetry accompanied by chiral symmetry. One might explain 
the observed rates using  heavy quark symmetry  for the strong 
vertices, while for the weak transition we rely on the  
existing knowledge of the form factors \cite{cheng}. The three-body 
decay with two vector meson states and one pseudoscalar is much more
difficult to approach.
 
The additional insight on the decay mechanism might come from the 
analysis of the $B$ meson two-body decays. Particularly 
interesting are the  decays $B^{\pm} \to  \phi K^{\pm}$, $B^{\pm} \to $ 
$\eta (\eta^{\prime}) K^{\pm}$ and $B \to K^{*\pm} \phi$. 
They have been extensively  studied using different existing 
techniques: the naive factorization \cite{heff2,Oh,BtoFKs},
the QCD factorization \cite{QCDF} and the SU(3) symmetry
\cite{Rosner}. Each of these decay modes is rather difficult to 
explain theoretically. 
The decays $B^{\pm} \to  \phi K^{\pm}$ and 
$B^{\pm} \to K^{*\pm} \phi$ might have significant annihilation 
contribution \cite{heff2,BtoFKs,dariescu}, but it  is not simple  
to have a consistent treatment of it. There is an interesting proposal
\cite{BtoFKs} in which the angular distributions of the final 
outgoing particles can be used to estimate the magnitude of 
annihilation contribution to the amplitude. However, we have to wait
for the new experimental data to extract the size of the annihilation 
contribution. The $B^{\pm} \to \eta (\eta^{\prime}) K^{\pm}$ decay rate
has not been easy to explain. It accounts for the well-known problem 
of the 
$\eta-\eta^{\prime}$ 
mixing \cite{mix,kroll} as can be seen from a variety of  
approaches used for  $B^{\pm} \to \eta (\eta^{\prime}) K^{\pm}$ 
\cite{heff2,neubert,heff1,heff3}. 
In the $B^{\pm} \to \eta (\eta^{\prime}) K^{\pm}$ decay 
mode,  it seems that the annihilation contribution is not very significant
\cite{heff2,heff1}.

One has to expect that the above described difficulties in 
these decay 
modes might appear in the three-body decay we discuss. Based on the 
current knowledge of two-body  transitions, we build a simple model 
which might clarify the role of the non-charm contributions in the 
${\rm BR}(B^{\pm} \to \phi \phi K^{\pm})$ decay. In our study of the 
$B^{\pm} \to \phi \phi K^{\pm}$ decay mechanism,  we follow the 
assumption in ref. \cite{cheng} and use double and single pole form factors 
for the $B$ meson semileptonic transitions \cite{becFF,FF}.
Our approach is based on  naive factorization, as
QCD factorization has not been developed yet for  three-body decays. 
The $SU(3)$ symmetry approach is not applicable due to the limited 
number of the observed $B$ decay modes. In our model 
we keep only dominant contributions
and as in the case of two-body charmless $B$ decays, we do not include 
annihilation contributions. 
We use a pole model including  the low-lying meson resonances 
and possible contributions  coming from higher mass excited states.

In order to compare our result with the Belle's result, 
we include in our calculation the interference 
between the non-resonant $B^- \to K^- \phi \phi$
and the resonant $B^- \to K^- \eta_c \to K^- \phi \phi$ decay 
amplitude.

In Section II we present the basic elements of our model, while in 
 Section III we {\bf give} the results for the three-body decay 
amplitude and discuss possible contributions to the decay rate. 

\vspace{.5cm}

{\bf II. THE MODEL}

\vspace{.5cm}

 The $\bar ub \to s\bar u \bar ss$  
transition which can produce two $ \phi$ mesons in the final state
via strong interactions,  can be realized by  the effective weak 
Hamiltonian \cite{Pham}-\cite{heff6}:

\begin{equation}
{\cal H}_{eff}=\frac{G_F}{\sqrt{2}}\left([V_{ub}V^*_{us}(c_1O_{1u}+c_2O_{2u})-
\sum^{10}_{i=3}[(V_{ub}V^*_{us}c^u_i+V_{cb}V^*_{cs}c^c_i+V_{tb}V^*_{ts}c^t_i)O_i]\right),
\label{heff}
\end{equation}
where $O_1$ and $O_2$ are the tree-level operators, $O_3-O_6$ are 
gluonic penguin operators and $O_7-O_{10}$ are electroweak penguin 
operators. Superscripts $u,c,t$ on the Wilson coefficients denote the 
internal quark in penguin loop. 
In order to apply the  
factorization approximation we rearrange the above operators 
using  
Fierz transformations and leave only color-singlet ones. One then comes 
to the effective weak Hamiltonian given by Eq. (\ref{heff}) 
replacing the coefficients $c_i$ by $a_i$. The relevant 
operators are:

\begin{equation}
\begin{array}{l}
{\cal O}_1=(\bar u b)_{V-A}(\bar su)_{V-A},\\
{\cal O}_2=(\bar s b)_{V-A}(\bar uu)_{V-A},\\
{\cal O}_3=\sum_q {\cal O}^q_3
=\sum_q (\bar s b)_{V-A}(\bar qq)_{V-A},\\
{\cal O}_4=\sum_q {\cal O}^q_4
=\sum_q (\bar s q)_{V-A}(\bar qb)_{V-A}, \\
{\cal O}_5=\sum_q {\cal O}^q_5=
\sum_q (\bar s b)_{V-A}(\bar qq)_{V+A},\\ 
{\cal O}_6=-2\sum_q {\cal O}^q_6
=-2\sum_q (\bar s(1+\gamma_5) q)(\bar q(1-\gamma_5)b), \\
{\cal O}_7=\sum_q {\cal O}^q_7
=\sum_q \frac{3}{2}e_q (\bar s b)_{V-A}(\bar qq)_{V+A},\\
{\cal O}_8=-2\sum_q {\cal O}^q_8=-2\sum_q \frac{3}{2}e_q
(\bar s(1+\gamma_5) q)(\bar q(1-\gamma_5)b), \\
{\cal O}_9=\sum_q {\cal O}^q_9
=\sum_q \frac{3}{2}e_q(\bar s b)_{V-A}(\bar qq)_{V-A}, \\
{\cal O}_{10}=\sum_q {\cal O}^q_{10}
=\sum_q \frac{3}{2}e_q(\bar s q)_{V-A}(\bar qb)_{V-A}, 
\end{array}
\end{equation}

The Wilson coefficients are taken from \cite{heff5}:
\begin{equation}
\begin{array}{lll}
a_1=1.05, & a_2=0.07, & a_3=47\times 10^{-4}, \\
a_4=(-43-16i)\times 10^{-3}, & a_5=-53\times 10^{-4}, 
& a_6=(-54-16i)\times 10^{-3}, \\
a_7=(0.4-0.9i)\times 10^{-4}, & a_8=(3.3-0.3i)\times 10^{-4}, 
& a_9=(-91-0.9i)\times 10^{-4}, \\
& a_{10}=(-13-0.3i)\times 10^{-4}.  & \\
\end{array}
\end{equation}

For the CKM matrix elements ($V_{ij}$) we  use  Wolfenstein 
parametrization:$V_{tb}V^*_{ts}=-A\lambda^2$ and
$V_{ub}V^*_{us}=A\lambda^4 (\bar\rho-i \bar \eta)$, where
$A=0.83$, $\lambda=0.222$, $\bar \rho=0.217/(1-\lambda^2/2)$ and
$\bar \eta=0.331/(1-\lambda^2/2)$.
The standard decomposition of the weak current matrix elements is: 
$$
\langle V(k,\varepsilon,m_V)|\bar q\Gamma^\mu
q|P(p,M)\rangle=\epsilon^{\mu\nu\alpha\beta}\varepsilon_\nu
p_\alpha k_\beta \frac{2V(q^2)}{M+m_V}+2im_V\frac{\varepsilon \cdot
q}{q^2}q^\mu A_0(q^2)
$$
\begin{equation}
+i(M+m_V)\Big[\varepsilon^\mu-\frac{\varepsilon
\cdot
q}{q^2}q^\mu\Big]A_1(q^2)-i\frac{\varepsilon\cdot q}{M+m_V}\Big[P^\mu-\frac{M^2-m_V^2}{q^2}q^\mu\Big]
A_2(q^2)\;,
\label{PvV}
\end{equation}
\\
\begin{equation}
\langle P(k,m_P)|\bar q\Gamma^\mu q|P(p,M)\rangle=
\left(P^\mu-\frac{ (M^2-m_P^2)}{q^2}q^\mu\right)F_+(q^2)+
\frac{(M^2-m_P^2)}{q^2} q^\mu F_0(q^2)\;,
\end{equation}
where $q^\mu=p^\mu-k^\mu$ and $P^\mu=p^\mu+k^\mu$. Also

\begin{equation}
\langle P(p)|\bar q\gamma^\mu(1-\gamma_5)q|0\rangle=if_P p^\mu\;, 
\qquad 
\langle 0|\bar q\gamma^\mu q|V(p)\rangle=g_V \varepsilon^\mu\;.
\label{fconst}
\end{equation}
Using experimental data \cite{PDG}, the  decay constants
are found to be $|g_\phi|=0.24\,$GeV$^2$, 
$|g_K|=0.19\,$GeV$^2$, $f_K=0.16\,$GeV and $f_\pi=0.132\,$GeV.
The lattice calculation \cite{lattice} gives for the $B$ 
meson decay constants $f_B=0.173\,$GeV and $f_{B_s}=1.22 f_B$. 
We also take $g_{B^*}=M_{B^*}f_B$ \cite{FF}. 
The $q^2$ dependence of the form factors is studied in  
\cite{FF}, where  a quark model is 
combined with a fit to lattice and experimental data. This approach 
results in  a double pole $q^2$ dependence  
of  $F_+(q^2)$, $V(q^2)$
and $A_0(q^2)$ 
\begin{equation}
f(q^2)=\frac{f(0)}{(1-q^2/M^2)(1-\sigma_1 q^2/M^2+\sigma_2 q^4/M^4)}\;,
\label{double}
\end{equation}
while for $A_{1,2}(q^2)$ and $F_0(q^2)$ \cite{FF} we have:
\begin{equation}
f(q^2)=\frac{f(0)}{(1-\sigma q^2/M^2+\sigma_2 q^4/M^4)}\;.
\label{single}
\end{equation}
Values of $M$, $f(0)$,
$\sigma_1$ and $\sigma_2$ are listed in Table 1.
  
\begin{table}
\begin{center}
\begin{tabular}{|c|c|c|c|c|c|c|}
\hline
form factor & $F_+$ & $F_0$ & $V$   & $A_0$ & $A_1$ & $A_2$  \\ \hline 
$f(0)$      & 0.36  & 0.36  & 0.44  & 0.45  & 0.36  & 0.32   \\ \hline
$\sigma_1$  & 0.43  & 0.70  & 0.45  & 0.46  & 0.64  & 1.23   \\ \hline
$\sigma_2$  & 0.0   & 0.27  & 0.0   & 0.0   & 0.36  & 0.38   \\ \hline
$M$ [GeV]   & 5.42  & 5.42  & 5.42   & 5.37  & 5.42  & 5.42   \\ \hline
\end{tabular}
\label{tabela}
\caption{The $B \to K,K^*$ form factors at $q^2 =0$ and the pole
parameters  \cite{FF}.}
\end{center}
\end{table} 

In the evaluation of  ${\cal O}_6$ operator we have as usual 
\cite{heff1}
\begin{eqnarray}
\bar q_1 \gamma_5 q_2 = \frac{-i}{m_1+m_2}
\partial_\mu (\bar q_1 \gamma^\mu \gamma_5 q_2)\,, \\
\bar q_1 q_2 = \frac{-i}{m_1-m_2}\partial_\mu 
(\bar q_1 \gamma^\mu q_2)\,.
\label{dirac}
\end{eqnarray}

The effects of strong interactions of light mesons are taken into 
account by using  the following effective Lagrangian  
\cite{chi1,chi2,chi3}:

\begin{equation}
{\cal L}_{\rm strong}=\frac{ig_{\rho\pi\pi}}{{\sqrt 2}}
Tr(\rho^\mu[\Pi,\partial_\mu \Pi])-
4\frac{C_{VV\Pi}}{f}\epsilon^{\mu\nu\alpha\beta} 
Tr(\partial_\mu \rho_\nu \partial_\alpha \rho_\beta \Pi)\;,
\label{strong}
\end{equation}
where $\Pi$ and $\rho^\mu$ are $3 \times 3$ matrices
containing pseudoscalar and vector meson field
operators respectively  and $f$ is a  pseudoscalar meson decay  
constant as in Eq. (\ref{fconst}).
We take $C_{VV \Pi}=0.31$ \cite{fajfer}.
In order to include  SU(3) flavor symmetry breaking, 
instead of the coupling constant coming from the $\rho \to \pi \pi$
decay ($g_{\rho\pi\pi} =5.9$), we use the coupling 
constant coming from the $\phi \to K K$ decay rate.
Thus, we have $g_{\phi K K}=6.4$. 

For the description of strong interactions between heavy and
light mesons, we use definitions given in 
\cite{FF} and heavy quark effective theory to get:
\begin{eqnarray}
<K (p_1) B^*_s (p_2,\varepsilon)|B(p_1+p_2)>=
\frac{g_{B_s^*BK}}{2}(p_1+2p_2)_\mu \varepsilon^\mu\,,\\
<\phi(p_1,\varepsilon_1)B^*_s(p_2,\varepsilon_2)|
B^*_s(p_1+p_2,\varepsilon)>=
\frac{1}{2}g_{B_sB_s\phi}( p_1\cdot\varepsilon_2
\,\varepsilon_1 \cdot \varepsilon-
p_1 \cdot \varepsilon \,
\varepsilon_2 \cdot \varepsilon_1)\,,
\end{eqnarray}
where $g_{B_sB_s\phi}f_{B_s}/2m_\phi=1.5\pm0.1$ and
$g_{B^*_sBK}f_{B_s}/2m_{B_s^*}=0.65\pm0.05$ \cite{FF}.

To  account for the $\eta - \eta^\prime$ mixing, we 
follow the approach in \cite{kroll}. Using the  quark basis 
($\eta_q \simeq (u \bar u + d \bar d)/{\sqrt 2}$ 
and $\eta_s \simeq s \bar s$), the mixing is given by
\begin{equation}
{\eta \choose \eta^\prime}=\left(
\begin{array}{cc} \cos\phi & - \sin\phi \\
\sin\phi & \cos\phi\end{array} \right) {\eta_q \choose \eta_s}\,,
\end{equation}
with the mixing angle $\phi=39.3^\circ \pm 1.0^\circ$.
The $\eta$, $\eta'$ decay constants are defined by 
\begin{equation}
<\eta|\bar q \gamma^\mu(1-\gamma_5)q|0>=if^q_\eta\,,
\qquad
<\eta^\prime|\bar q \gamma^\mu(1-\gamma_5)q|0>=if^q_{\eta^\prime}\,,
\end{equation}
where 
\begin{equation}
\begin{array}{ll}
f^{u,d}_\eta=f_{u,d}\cos\phi /\sqrt{2}\,, &
f^s_{\eta}=-f_s \sin\phi\,, \\
f^{u,d}_{\eta^\prime}=f_{u,d}\sin\phi /\sqrt{2}\,, &
f^s_{\eta^\prime}=f_s\cos\phi\,, 
\end{array}
\end{equation}
with $f_{u,d}=(1.07 \pm 0.02)f_\pi$ and $f_s=(1.34 \pm 0.06)f_\pi$.
The form factors for the $B \to \eta (\eta')$ transition 
can be written as:
\begin{equation}
F^\eta_{0,+}(q^2)=F^\pi_0(q^2)\cos\phi/\sqrt 2\,, \qquad
F^{\eta^\prime}_{0,+}(q^2)=F^\pi_0(q^2)\sin\phi/\sqrt 2\,.
\end{equation}
The $q^2$ dependence of $F^\pi_0$ is described by Eq. (\ref{single}),
with $F^\pi_0(0)=0.29$, $\sigma_1=0.76$ and $\sigma_2=0.28$ while 
the $q^2$ dependence of $F^\pi_+$ is described by
Eq. (\ref{double}), with $F^\pi_+(0)=0.29$ and $\sigma_1=0.48$
\cite{FF}.

Before we consider  the  
$B^- \to \phi \phi K^-$ decay amplitude,  
we check how our model 
works for the    two-body 
decays: $B^- \to \eta  K^-$, $B^- \to \eta^\prime K^-$,
$B^- \to \phi K^-$,
and $B^- \to \phi K^{*-}$. Namely, $B^- \to \phi \phi K^-$ can occur 
through one of these decay chains:
$B^- \to \eta(\eta^\prime) K^-$ followed by $\eta(\eta^\prime) \to
\phi \phi$; $B^- \to \phi K^-$, followed by  $K^- \to K^- \phi$
and $B^- \to \phi K^{*-}$ followed by $K^{*-} \to K^- \phi$. 
These  decays have already been studied within the factorization 
approximation by Ali {\it et. al.} \cite{heff2}.
Using their formulas for the amplitudes 
with the Wilson coefficients, the form factors and other parameters 
{\bf as} given above, we obtain the  branching ratios 
for the two-body decays 
presented in Table 2 together with the  experimental results. 
We point out that references \cite{heff2,heff3}, 
as well as our predictions,  include
the axial anomaly contribution in  
$b \to s g g \to s \eta (\eta')$.
In our calculation,
the contribution of the $\bar cc$ component in $\eta, \eta^\prime$ 
was found 
to be small and therefore we safely neglect it. \\

\begin{table}
\begin{center}
\begin{tabular}{|c|c|c|c|}
\hline
 & Belle \cite{Belle-Fes,Belle-FKs}& BaBar 
 \cite{BaBar1,BaBar2,BaBar3,BaBar4} & our model
 \\ \hline 
$B^- \to \eta K^-$  &
$(5.3^{+1.8}_{-1.5}\pm0.6)\times 10^{-6}$ &
$(2.8^{+0.8}_{-0.7}\pm0.2)\times 10^{-6}$ &    
$ 2.1 \times 10^{-6}$  \\  \hline
$B^- \to \eta^\prime K^-$  & &
$(7.67\pm 0.35\pm 0.44)\times 10^{-5}$ & $ 3.0\times 10^{-5}$  \\ \hline
$B^- \to \phi K^-$ &
$(9.4\pm 1.1\pm 0.7)\times 10^{-6}$ & $(10^{+0.9}_{-0.8}\pm0.5)
\times 10^{-6}$ & 
$ 14.9\times 10^{-6}$  \\ \hline
$B^- \to \phi K^{*-}$  &
$(6.7^{+2.1+0.7}_{-1.5-0.8})\times 10^{-6}$  &  
$(12.7^{+2.2}_{-2.0}\pm1.1)\times 10^{-6}$   &
$ 8.6\times 10^{-6}$    \\ \hline
\end{tabular}
\label{tabela2}
\caption{The experimental and theoretical results  for the relevant 
$B^-$ two-body decay rates. 
The rates for $B^- \to \eta K^-$ and  
$B^- \to \eta^\prime K^-$ are calculated without 
$c \bar c$ contributions.}
\end{center}
\end{table} 

{\bf III.  THE $B^- \to \phi \phi K^-$ DECAYS}

\vspace{0.5cm}
 
\begin{figure}
\begin{center}
\includegraphics{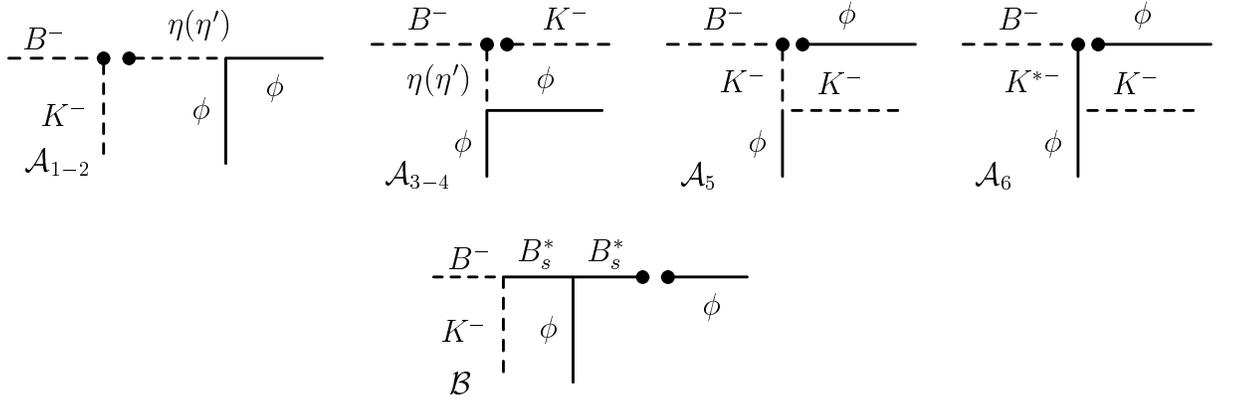}
\caption{Feynman diagrams for  $B^- \to \phi \phi K^-$.}
\end{center}
\end{figure}

{ The dominant contributions in the   $B^-(p) \to K^-(p_1) 
\phi(p_2) \phi(p_3)$ decay 
amplitude with the $\phi\phi$ invariant mass in the region below
the charm threshold are shown in Fig. 1.} We write the amplitude 
for this decay in the following form{\bf:}
\begin{equation}
{\cal M}={\cal A}_1 \left(\sum_{q=s,u,d}C^\eta_{1q}\right)+
{\cal A}_2 \left(\sum_{q=s,u,d}C^{\eta^\prime}_{1q}\right)+
{\cal A}_3 C^\eta_2+{\cal A}_4 C^{\eta^\prime}_2+
({\cal A}_5+{\cal A}_6+{\cal B})C^K_1\,.
\label{Amplituda}
\end{equation}
Here
$$
C^{\eta}_{1u}=\frac{G_F}{\sqrt 2}
\left(V_{ub}V^*_{us}a_2-V_{tb}V^*_{ts}(a_3-a_5+a_9-a_7)\right)
\frac{f^u_\eta}{f_\pi}\,, $$$$
C^{\eta}_{1d}=-\frac{G_F}{\sqrt 2}
V_{tb}V^*_{ts}(a_3-a_5-1/2(a_9-a_7))\frac{f^d_\eta}{f_\pi}\,,
$$
\begin{equation}
C^{\eta}_{1s}=-\frac{G_F}{\sqrt 2}
V_{tb}V^*_{ts}\left(\left(a_3+a_4-a_5-1/2(a_9+a_{10}-a_7)+
\frac{p^2_\eta}{m_bm_s}(a_6-1/2a_8)\right)\frac{f^s_\eta}{f_\pi}-
\right.$$$$\left. \frac{p^2_\eta}{m_bm_s}(a_6-1/2a_8)
\frac{f^u_\eta}{f_\pi}\right)\,,
\end{equation}
$$
C^{\eta}_{2}=\frac{G_F}{\sqrt 2}
V_{ub}V^*_{us}a_1-V_{tb}V^*_{ts}(a_4+a_{10}+\frac{2p^2_K}{m_bm_s}
(a_6+a_8))\,,
$$
 \begin{equation}
C^{K}_{1}=\frac{G_F}{\sqrt 2}
(-V_{tb}V^*_{ts})\big(a_3+a_4+a_5-1/2(a_7+a_9+a_{10})
\big)\,, \qquad
\end{equation}
where $p_\eta$ and $p_K$ are the momenta of $\eta$ and $K$ 
meson respectively. The formulas for $\eta^\prime$ are obtained by 
replacing $f^q_{\eta}$ and $k_s$ with $f^q_{\eta^\prime}$
and $k^\prime_s$. The constant $k_s$ 
($k^\prime_s$) projects the $\bar ss$ component of the 
$\eta$ ($\eta^\prime$) meson and it is equal $-\sin\phi$ for $\eta$ 
and $\cos\phi$ for $\eta^\prime$. The coefficient $C_{1s}^{\eta}$ 
contains the effect of the axial anomaly as in \cite{heff2,heff3}. 
The amplitudes are determined by 
\begin{equation}
\label{A1}
{\cal A}_{1}=8iC_{VV\Pi}k_s F_0(x)\frac{M^2-m_K^2}{m^2_{\eta}-x} 
\epsilon_{\mu\nu\alpha\beta}
\varepsilon^\mu_2 \varepsilon^\nu_3 p_2^\alpha p_3^\beta\,, 
\end{equation}
\begin{equation}
\label{A2}
{\cal A}_{2}=8iC_{VV\Pi}k_s F_0(x)
\frac{M^2-m_K^2}{m^2_{\eta^{\prime}}-x}
\epsilon_{\mu\nu\alpha\beta}
\varepsilon^\mu_2 \varepsilon^\nu_3 p_2^\alpha p_3^\beta\,, 
\end{equation}
\begin{equation}
\label{A3}
{\cal A}_{3}=8iC_{VV\Pi}\frac{f_K}{f_\pi}
\big(F^{\eta}_0(m_K^2)\frac{M^2-m^2_{\eta}}
{m^2_{\eta}-x}+F^{\eta}_+(m_K^2)\big)
\epsilon_{\mu\nu\alpha\beta}
\varepsilon^\mu_2 \varepsilon^\nu_3 p_2^\alpha p_3^\beta\,, 
\end{equation}
\begin{equation}
\label{A4}
{\cal A}_{4}=8iC_{VV\Pi}\frac{f_K}{f_\pi}
\big(F^{\eta^{\prime}}_0(m_K^2)\frac{M^2-m^2_{\eta^{\prime}}}
{m^2_{\eta^{\prime}}-x}+F^{\eta^{\prime}}_+(m_K^2)\big)
\epsilon_{\mu\nu\alpha\beta}
\varepsilon^\mu_2 \varepsilon^\nu_3 p_2^\alpha p_3^\beta\,, 
\end{equation}
\begin{equation}
\label{A5}
{\cal A}_{5}=
2\sqrt 2 g_\phi g_{\rho\pi\pi}F_+(m_\phi^2)\left(
\frac{p_1 \cdot \varepsilon_2\, p \cdot \varepsilon_3}{m_K^2-y}
+\frac{p \cdot \varepsilon_2\, p_1 \cdot \varepsilon_3}
{x+y-M^2-2m_\phi^2}\right)\,,
\end{equation}
\begin{equation}
\label{A6}
{\cal A}_{6}=-\frac{2g_\phi C_{VV\Pi}}{f_K(M+m_{K^*})}
\left(2i(M+m_{K^*})^2A_1(m_\phi^2)
\left(\frac{\epsilon_{\mu\nu\alpha\beta}
\varepsilon^\mu_2\varepsilon^\nu_3p^\alpha_1p^\beta_2}
{m_{K^*}^2-y}+ \right.\right.
\end{equation}
$$\left.\left.\frac{\epsilon_{\mu\nu\alpha\beta}
\varepsilon^\mu_2\varepsilon^\nu_3p^\alpha_1p^\beta_3}
{M^2-2m_\phi^2+m_K^2-m^2_{K^*}-x-y}\right)\right. 
$$ $$
\left.-4iA_2(m_\phi^2)\left(\frac{
\epsilon_{\mu\nu\alpha\beta}\varepsilon^\mu_2
p^\nu_1p^\alpha_2p^\beta_3(p_1+p_2)\cdot\varepsilon_3}
{m^2_{K^*}-y}+\frac{\epsilon_{\mu\nu\alpha\beta}\varepsilon^\mu_3
p^\nu_1p^\alpha_2p^\beta_3(p_1+p_3)\cdot\varepsilon_2}
{M^2+2m_\phi^2+m_K^2-m^2_{K^*}-x-y}\right)\right. $$
$$\left.+V(m_\phi^2)\left(\left(
\frac{(M^2-m_\phi^2)(m_\phi^2-m_K^2)+y(M^2+2m_\phi^2+m_K^2-2x)-y^2}
{m_{K^*}^2-y}
\right.\right.\right.$$$$\left.\left.\left.
+\frac{M^2(m_\phi^2-m_K^2-x+y)-(x-y)(m_K^2-x-y)
+(m_K^2-2x-2y)m_\phi^2-m_\phi^4}{M^2+2m_\phi^2+m_K^2-m^2_{K^*}-x-y}
\right) \varepsilon_2 \cdot \varepsilon_3\right.\right.
$$$$ \left.\left.
-2\frac{M^2+3m_\phi^2-x-y}{x+y-M^2-2m_\phi^2-m_K^2+m^2_{K^*}}
p_1 \cdot \varepsilon_2\, p \cdot \varepsilon_3
-2\frac{m_\phi^2-m_K^2+y}{m^2_{K^*}-y}
p_1 \cdot \varepsilon_3\, p \cdot \varepsilon_2
\right.\right. $$$$ \left.\left.
2p_1 \cdot \varepsilon_2\, p_1 \cdot \varepsilon_3
\left(\frac{y-M^2-m_\phi^2}{M^2+2m_\phi^2+m_K^2-m^2_{K^*}-x-y}
+\frac{x+y-m_\phi^2-m_K^2}{m^2_{K^*}-y}\right)
\right.\right. $$$$ \left.\left.
+2(M^2+m_K^2-x)\left(
\frac{p_3 \cdot \varepsilon_2\, p_1 \cdot \varepsilon_3}
{M^2+2m_\phi^2+m_K^2-m^2_{K^*}-x-y}-
\frac{p_1 \cdot \varepsilon_2\, p_2 \cdot \varepsilon_3}
{m^2_{K^*}-y}
\right)
\right.\right. $$$$ \left.\left.
\frac{2(y-m_\phi^2+m_K^2)
p \cdot \varepsilon_2\, p_2 \cdot \varepsilon_3}{m^2_{K^*}-y}
+\frac{2(x+y-M^2-m_\phi^2-2m_K^2)
p_3 \cdot \varepsilon_2\, p \cdot \varepsilon_3}
{M^2+2m_\phi^2+m_K^2-m^2_{K^*}-x-y}\right)\right)\,,
$$
\begin{equation}
\label{B2}
{\cal B}=\frac{g_{BB\phi}g_{BBK}g_\phi f_{B_s^*}}
{4M_s^2(M_s^2-m_\phi^2)(M_s^2-x)}
\left(\varepsilon_2 \cdot \varepsilon_3
\left((M^2-m_K^2)x-M_s^2(M^2+4m_\phi^2-m_K^2)\right)
\right. \end{equation}$$ \left.
+(M_s^2-M^2+m_K^2)(p_3 \cdot \varepsilon_2 \,p \cdot \varepsilon_3+
p \cdot \varepsilon_2 \,p_2 \cdot \varepsilon_3)
+(M_s^2+M^2-m_K^2)(p_3 \cdot \varepsilon_2 \,p_1 \cdot \varepsilon_3+
p_1 \cdot \varepsilon_2 \,p_2 \cdot \varepsilon_3)\right)\,.$$

In our expressions the two $\phi$ meson polarization vectors 
are denoted by $\varepsilon_2 \equiv \varepsilon_2(p_2)$ and 
$\varepsilon_3 \equiv \varepsilon_3(p_3)$, the 
$B^-$ and $B^{*-}_s$ masses are $M$ and $M_s$, and $m_N$ stands 
for the mass of  $N$ meson. 

To obtain  the decay width, we make the following 
integration over the Dalitz plot:

\begin{equation}
\Gamma=  \frac{1}{2}\frac{1}{(2\pi)^3}\frac{1}{32M^3}
\int {|{\cal M}|^2}dxdy\,, 
\label{Br}
\end{equation}
where $y=m^2_{12}=(p_1+p_2)^2$, $x=m^2_{23}=(p_2+p_3)^2$. 
Note that we include the factor  $1/2$ 
due to two
identical mesons in the final state. In the above integral,
upper and lower bounds for $y$ are:
\begin{eqnarray}
y_{max}=(E^*_1+E^*_3)^2-\left(\sqrt{E^{*2}_1-m^2_K}-
\sqrt{E^{*2}_3-m^2_\phi}\right)^2\,,
\\
y_{min}=(E^*_1+E^*_3)^2-\left(\sqrt{E^{*2}_1-m^2_K}+
\sqrt{E^{*2}_3-m^2_\phi}\right)^2\,,
\end{eqnarray}
with {\bf the} energies $E^*_1$ and $E^*_3$ given by:
\begin{eqnarray}
E^*_1=\sqrt{x}/2\,, \qquad
E^*_3=(M^2-x-m_\phi^2)/(2\sqrt{x})\,. 
\end{eqnarray}
\noindent
The integration over $x$ is bounded by $x_{min}=4m_\phi^2$ and
$x_{max}=(M-m_K)^2$.

First we consider only the phase space region  with the $\phi\phi$
invariant mass below the $\eta_c$
threshold by taking  $x<(2.85\,$GeV$)^2$.
The Belle collaboration has measured 
${\rm BR}(B^- \to K^- \phi \phi)_{x<(2.85\,{\rm GeV})^2} $ $=
(2.6^{+1.1}_{-0.9}\pm0.3) \times 10^{-6}$ while
our model  gives
${\rm BR}(B^- \to K^- \phi \phi)_{x<(2.85\,{\rm GeV})^2}=1.8
\times 10^{-6}\,.$ 
The calculated decay rate is the total contributions from the
parity violating (the  terms in amplitude containing  
$\epsilon_{\mu \nu \alpha \beta}$) and parity conserving parts, 
which do not interfere. 
The parity violating component gives the rate $1.5 \times 10^{-6}$, 
while from the parity conserving part we get 
$0.3 \times 10^{-6}$. 
{ We note that  the dominant contribution comes from the 
$\eta$, $\eta^\prime$ intermediate states in  the graph 
${\cal A}_{1-2}$ of Fig. 1 
and its contribution alone gives the branching ratio of $1.3 \times 10^{-6}$.  
Since for the  $B \to \eta (\eta^\prime) K$ decay rates
the annihilation term is not very large \cite{heff2}, we do not expect a significant change in the  
$B \to \phi \phi K$ decay rate if its effects 
are taken into account.}

In addition to the low-lying mesons such as $\eta$ and $\eta^\prime$,
one could expect that  higher mass excited
states in the  $1-2\,$ GeV  region could  also make important 
contribution to the amplitude. 
If the $\eta, \eta^\prime$  in the diagram,
${\cal A}_{1-2}$ (Fig. 1) are replaced by scalar or tensor mesons, which contain  
$\bar s s$ (e.g. $f_0(980)$, $f_2(1270)$) one finds that both 
contributions are suppressed. The  observed
rate $B^- \to f_0(980)K^-$ \cite{fry} is by an order 
of magnitude smaller than the rate of  $B \to \eta' K^-$ 
and the decays of $B$ into a pseudoscalar and a tensor meson, 
are expected to have branching rations of the order  $10^{-8}$
\cite{Kim}.
The products 
$<f_{0,2}|\bar (\bar u b)_{V\pm A,S\pm P}|B^-><K^-|\bar (\bar s u)_{V\pm A,S\pm P}|0>$,
($V\pm A$ stands for the left and right handed currents, and $S\pm P$ are 
scalar and pseudoscalar densities) 
can be safely neglected because of the small values of the $B \to S, T$
 transition form factor involved in the  graphs like those in 
${\cal A}_{3-4}$ (Fig. 1) \cite{Kim1}.
The same arguments hold for the higher mass $\bar us$ excited
states. 

However, a large contributions to the decay rate can be expected from
the higher mass excited states with the quantum numbers of
$\eta, \eta^\prime$: $\eta(1260)$, $\eta(1490)$ \cite{asimovic}. In the study of \cite{Castro}
it has been found that $\eta(1295)$ is most likely $(\bar u u + \bar
dd)/\sqrt{2}$, while $\eta(1490)$ is almost pure ${\bar ss}$
state. Therefore, we might expect the presence of $\eta(1490)$ in the
diagram ${\cal A}_{1-2}$. Unfortunately, its interactions are very 
poorly known and we can make only a very rough  
 estimation of the 
$\eta(1490) \phi \phi$ coupling within a naive quark model. 
The coupling of the
$\eta$ or any state with the same quantum numbers like $\eta(1490)$ 
$\to \phi \phi$ could be estimated
by a $s\bar s$ quark loop triangle graph as shown in Fig. 2.

\begin{figure}
\begin{center}
\includegraphics[width=6cm]{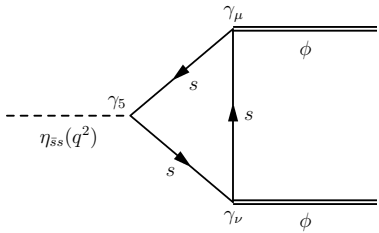}
\caption{$\eta \phi \phi$ interaction.} 
\end{center}
\end{figure}
 We then find that

\begin{equation}
C_{\eta \phi\phi}(q^2) \propto
\int^1_0 dx \int^{1-x}_0 dy
(m_s^2-xm_\phi^2-ym_\phi^2+x^2m_\phi^2+y^2m_\phi^2-xy(q^2-2m_\phi^2))^{-1}\,.
\end{equation}
Taking the dynamical s quark mass $m_s \sim 500\,$MeV, we 
roughly estimate 
$C_{\eta\phi\phi}(m_{\eta}^2):C_{\eta^\prime\phi\phi}(m_{\eta^{\prime}}^2):
C_{\eta(1490)\phi\phi}(m_{\eta(1490)}^2)=1:0.85:0.40$.
We fix $C_{\eta\phi\phi}(m_{\eta^2})$ to be equal to the 
vector-vector-pseudoscalar coupling
$C_{VVP}=0.31$.  

Including the contribution of  $\eta(1490)$ in the graph 
${\cal A}_{1-2}$ with above couplings, and assuming that there is no
axial anomaly term in the coefficient $C_{1s}^{\eta}$, we find:
\begin{equation}
{\rm BR}(B^- \to K^- \phi \phi)_{x<(2.85\,{\rm GeV})^2}= 3.7  
\times 10^{-6}\,.
\end{equation} 
\begin{figure}
\begin{center}
\includegraphics{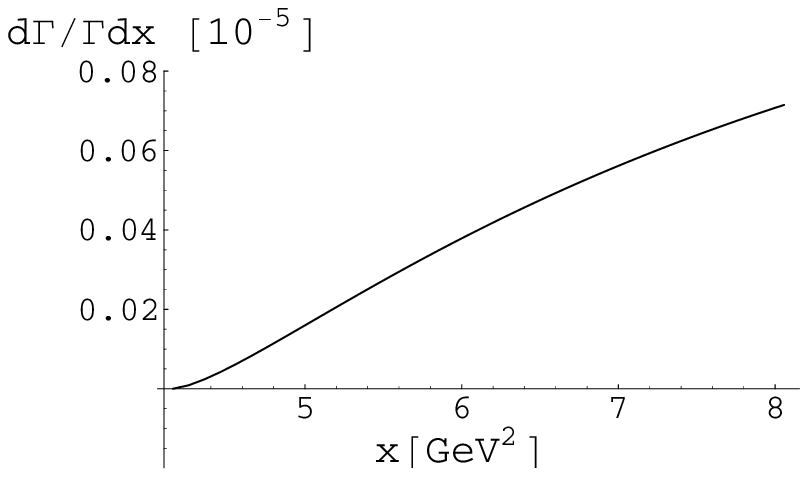}
\includegraphics{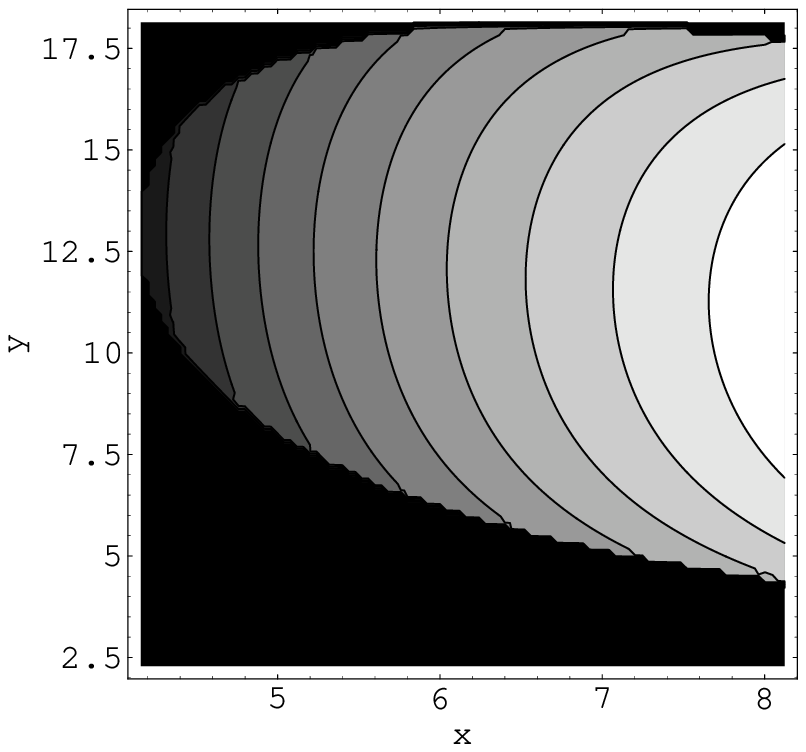}
\caption{$\frac{1}{\Gamma}\frac{d\Gamma}{dx}$
spectrum for the $B^- \to K^- \phi \phi$ decay  
with the $\phi\phi$ invariant mass in the region below 
charm threshold  and the Dalitz plot.} 
\end{center}
\end{figure}
The distribution $(1/\Gamma)d\Gamma/dx$ 
as the function of the $\phi\phi$ invariant mass in 
the region below 
charm threshold and the
Dalitz plot are is given in Fig. 2 for this case.

Note that the non-resonant contribution in the branching ratio, 
 measured by Belle collaboration 
contains not only the non-resonant amplitude itself, but also the 
interference terms with the resonant contribution 
$B^- \to K^- \eta_c \to K^- \phi \phi$ as in \cite{BFOP}.
In addition to the $\eta_c$ state there are 
a number of other $c \bar c$ bound states which might contribute.
From these, the biggest contribution will  arise  from
the $\chi_{c0}$ state  as its mass is closest to the region we discuss 
($x<2.85\,{\rm GeV}$). This contribution can be obtained from the measured
$B^- \to \chi_{c0} K^-$ decay rate \cite{chi0}.
 One might then expect
that the $B^- \to \chi_{c0} K^-  \to $ $ \phi \phi K^- $ 
transition 
can give additional 
interference with the calculated rates. However, the  rate 
for $\chi_{c0} \to \phi \phi$ is ten times smaller than the rate 
$\eta_c \to  \phi \phi$ and we expect additional suppression.
This leads us to the conclusion that the interference of the 
non-resonant and  the resonant terms from the $c \bar c$ states 
other than $\eta_c $ is negligible below the charm threshold. 
 
Next, we comment on the interference of the 
$\eta_c$ resonance with the non-resonant contribution
in the region of the phase space with the invariant mass of the
$\phi \phi$ state within the region
$(2.94\,{\rm GeV})^2<x<(3.02\,{\rm GeV})^2$.
The decay rate for $B^- \to K^- \eta_c$ 
is not theoretically very well understood. Naive factorization 
leads to a decay rate  ten times
smaller  than the branching ratios $6.9^{+3.4}_{-3.0} \times 10^{-4}$ measured by 
CLEO collaboration \cite{BvEcCLEO}, 
$(1.34\pm 0.09\pm0.13\pm0.41) \times 10^{-3}$ by BaBar collaboration 
\cite{BvEcBaBar} or 
$(1.25 \pm 0.14^{+0.10}_{-0.12}\pm 0.38) \times 10^{-3}$ 
by Belle  collaboration \cite{BvEcBelle}. QCD factorization seems 
to face similar problem in explaining 
this decay amplitude
 \cite{song}.
On the other hand, the decay $\eta_c \to  \phi \phi$ rate 
is  not very well
understood. First, the statistics for the 
$\eta_c \to  \phi \phi$ decay rate is rather poor (the 
error stated in \cite{PDG}
seems to be underestimated \cite{zivko}). Secondly, by assuming the
SU(3) flavor symmetry one cannot reproduce both the
$\eta_c \to  \phi \phi$ and the $\eta_c \to  \rho \rho$ measured decay rates.

Facing these difficulties 
we use experimental data to  
estimate the size of this resonant contribution. 
In the phase space region $(2.94\,{\rm GeV})^2<x<(3.02\,{\rm GeV})^2$, 
Belle measures
${\rm BR}(B^- \to K^- \eta_c)\times {\rm BR}(\eta_c \to \phi \phi)=
(2.2^{+1.0}_{-0.7}\pm 0.5)\times 10^{-6}$.
We model ${\rm BR}(B^- \to K^- \eta,\eta^\prime)\times {\rm BR}(\eta,\eta^\prime 
\to \phi \phi)$ by taking the   $\eta_c$ propagator and 
by fitting the Belle's data.
The results for the interference are given on Fig. 3, where 
we present both cases:  positive and negative interference terms, which is 
the result of an  unknown phase in the $\eta_c \to  \phi \phi$ 
decay amplitude.

\begin{figure}
\begin{center}
\includegraphics[width=10cm]{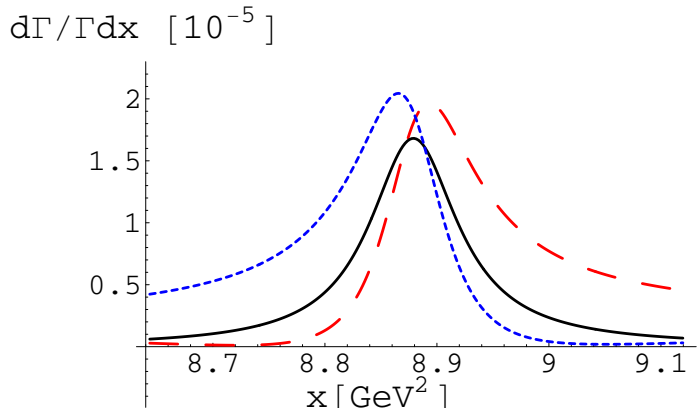}
\caption{The $\frac{1}{\Gamma}\frac{d\Gamma}{dx}$
spectrum for $B^- \to K^- \phi \phi$ decay in $\eta_c$ 
resonance region. In full (black) line, we show only the resonant
contribution while dotted (blue) and dashed (red) show the destructive
and constructive interference with the non charm amplitude respectively.} 
\end{center}
\end{figure}

The contribution of the $\eta_c$ resonance in the 
$x<(2.85\,{\rm GeV})^2$  
region can affect the non-resonant
  branching ratio,  
reducing it to $3.3\times 10^{-6}$, in the case of destructive
interference, or increasing it to  $4.2\times 10^{-6}$ in the opposite
case.

 In the treatment of the decay 
$B^- \to  \phi \phi K^-$ , due to  the complexity
of the problem, there are uncertainties which might be important. 
The  
simplest possible approach  which  will give us a reasonable estimate
of the decay rates could be the use of factorization model 
for the weak vertices  and the creation of the final state by the 
exchange of resonant states. 
Both assumptions bring in uncertainties themselves. The model should be 
tested  when more experimental data  
 on  other $B$ decays into two vector and one pseudoscalar 
 states will  be available. 
The additional errors come from: the lack of understanding of 
the $B^- \to K^- \eta'$ 
decay amplitude within the factorization approximation;
the treatment of the two gluon exchange in the amplitudes of the 
$\eta K$, $\eta' K$ modes \cite{kreso} 
and the assumptions on $B^- \to \eta(1490) K^-$ decay mechanism.  
The other input parameters might introduce about $10\%$ uncertainty.
Since the 
$\eta(1490)$ state gives 
important contribution to the rate, the  theoretical ignorance 
of its  coupling is potentially dangerous.

In conclusion, we have constructed a model,  based on the naive 
factorization and the exchange of intermediate resonances, 
with the aim to understand the decay mechanism in the 
$B^- \to \phi \phi K^-$ 
decay  with the $\phi\phi$ invariant mass in the region below 
charm threshold. 
 We have found that the largest contribution in the rate comes from the decay chain
 $B^- \to \eta (\eta', \eta(1490) ) K^- \to \phi \phi K^-$. 
Although this dominant contribution comes from the tree-level and 
penguin operators, we find that effects of the tree 
amplitudes are negligible. The interference effects  of the 
$\eta_c$ resonance with the non-resonant contribution
in the region of the phase space with the invariant mass of the
$\phi \phi$ state in the region
$(2.94\,{\rm GeV})^2<x<(3.02\,{\rm GeV})^2$ might decrease 
 (or increase) the rate by  $\sim 10\%$, depending on the sign of the 
interference term.

\vspace{0.5cm}
{\bf ACKNOWLEDGMENTS}
\vspace{0.5cm}
 
We thank our colleagues P. Kri\v zan, B. Golob and T. \v Zivko 
for stimulating discussions on experimental aspects
of this investigation. The research of S. F. and A. P. was
supported in part by the Ministry of Education, Science and
Sport of the Republic of Slovenia.

\end{document}